# THE HOST GALAXIES OF $Z \approx 2$ RADIO-QUIET QSOS

ITZIAR ARETXAGA, B.J. BOYLE AND R.J. TERLEVICH
*Royal Greenwich Observatory*
*Madingley Road. Cambridge CB3 0EZ. UK.*

Based on an $R$-band imaging survey of high-redshift ($z \approx 2$) QSOs with the 4.2m WHT, we report the detection of extensions to the nuclear PSFs of two radio-quiet and one radio-loud QSO. The extensions are most likely host galaxies, with luminosities of at least $3 - 7\%$ of the QSO luminosity. The most likely values for the luminosities lie in the range $6 - 18\%$ of the QSO luminosity ($R \sim 19.8 - 20.9$ mag). Our observations show that, if the extensions we have detected are indeed galaxies, extraordinary massive and luminous galaxies are not only characteristic of radio-loud objects, but of QSOs as an entire class. For a detailed description of this work, see Aretxaga, Boyle & Terlevich (1995) MNRAS, 275, L27.

*Figure 1:* The left panel shows a $50'' \times 50''$ field of one of the radio-quiet QSOs (1630.5+3749) for which we have found and extension to its nuclear PSF. The right pannel shows the field after subtracting a 2D PSF profile to both QSO and stars. Note that the stars are subtracted while the QSO is left with significant residuals ($> 3\sigma$).

IA's work is supported by the EEC HCM fellowship ERBCHBICT941023. She acknowledges the organizers for partial support to attend the meeting.